\documentclass[twocolumn]{autart}

\usepackage{amssymb,amsmath,stackrel,amsopn,latexsym}
\usepackage{graphicx}
\usepackage{epstopdf}
\usepackage{amsfonts}
\usepackage[authoryear]{natbib}
\usepackage{color}

\usepackage{enumitem}

\graphicspath{{img/}}

\newcommand*{\red}{\textcolor{black}}

\newcommand{\eio}{e^{i\omega}}

\newcommand{\R}{\mathbb{R}}

\newcommand{\M}{\mathcal{M}}

\newtheorem{definition}{Definition}
\newtheorem{proposition}{Proposition}
\newtheorem{theorem}{Theorem}
\newtheorem{example}{Example}
\newtheorem{corollary}{Corollary}
\newtheorem{remark}{Remark}
\newtheorem{lemma}{Lemma}
\newtheorem{assumption}{Assumption}

\begin{document}

\begin{frontmatter}
\title{Identifiability of linear dynamic networks\thanksref{footnoteinfo}
}

\thanks[footnoteinfo]{Third revision 28 September 2017. Earlier versions of this paper were submitted on 26 August 2016, 6 February 2017 and 28 June 2017. This project has received funding from the European Research Council (ERC), Advanced Research Grant SYSDYNET, under the European Union's Horizon 2020 research and innovation programme (grant agreement No 694504).}

\author[First]{Harm H.M. Weerts},
\author[First]{Paul M.J. Van den Hof} and
\author[Second]{Arne G. Dankers}

\address[First]{Control Systems Group, Department of Electrical Engineering, Eindhoven University of Technology, The Netherlands (email: h.h.m.weerts@tue.nl, p.m.j.vandenhof@tue.nl)}
\address[Second]{Department of Electrical Engineering, University of Calgary, Canada (email: adankers@hifieng.com)}


\begin{abstract}
Dynamic networks are structured interconnections of dynamical systems (modules) driven by external excitation and disturbance signals. In order to identify their dynamical properties and/or their topology consistently from measured data, we need to make sure that the network model set is identifiable. We introduce the notion of \textit{network identifiability}, as a property of a parameterized model set, that ensures that different network models can be distinguished from each other when performing identification on the basis of measured data. Different from the classical notion of (parameter) identifiability, we focus on the distinction between network models in terms of their transfer functions.
For a given structured model set with a pre-chosen topology, identifiability typically requires conditions on the presence and location of excitation signals, and on presence, location and correlation of disturbance signals. Because in a dynamic network, disturbances cannot always be considered to be of full-rank, the reduced-rank situation is also covered, meaning that the number of driving white noise processes can be strictly less than the number of disturbance variables. This includes the situation of having noise-free nodes.
\end{abstract}
\end{frontmatter}

\section{Introduction}
\label{Sec:Introduction}
\begin{figure}[hbt]
		\centering
			\includegraphics[width=1\columnwidth]{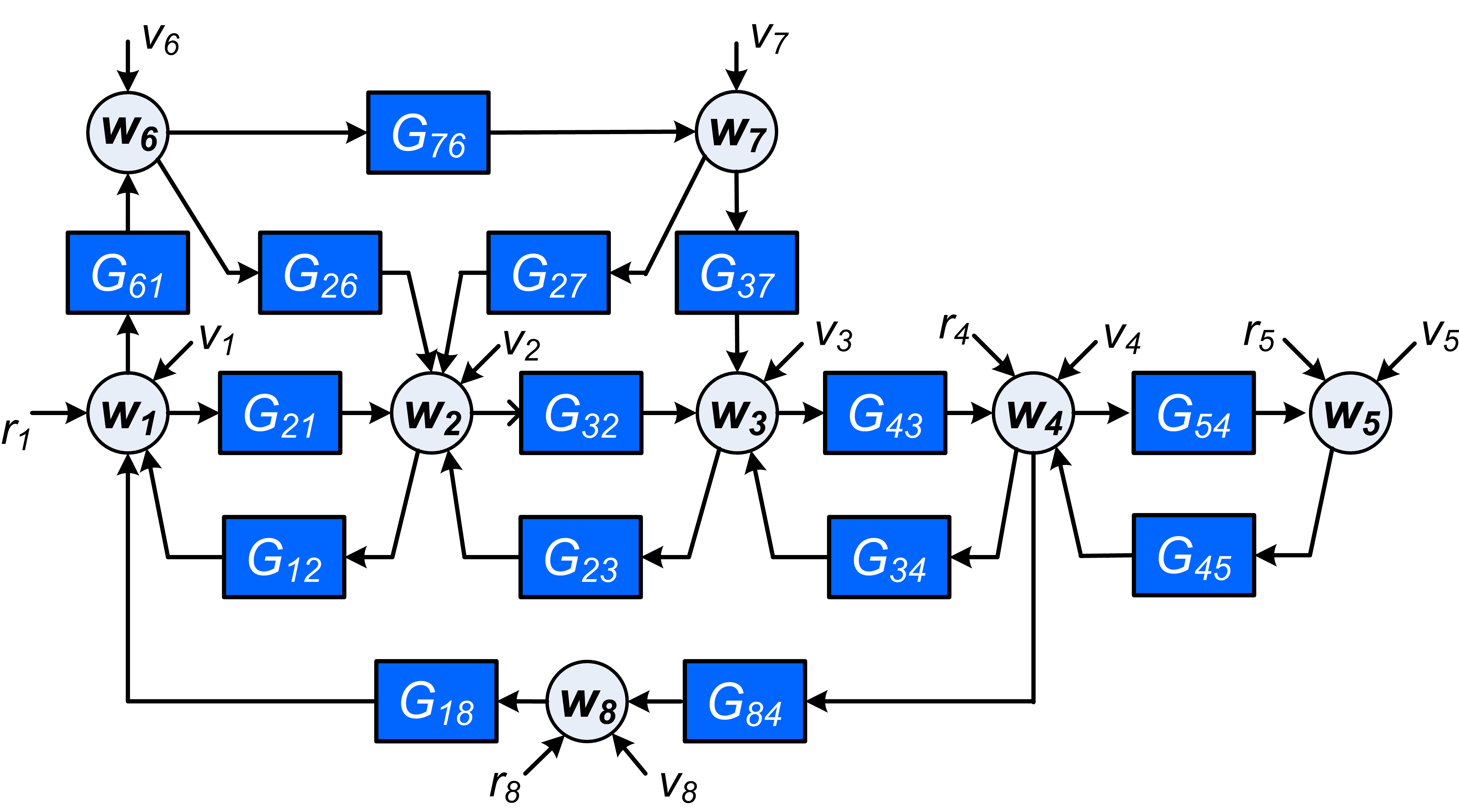}
		\caption{Dynamic network \red{where} node variables $w_i$ are the outputs of the summation points indicated by circles.}
		\label{fig:examnetw}
\end{figure}

Dynamic networks are structured interconnections of dynamic systems and they appear in many different areas of science and engineering. Because of the spatial connections of systems, as well as a trend to enlarge the scope of control and optimization, interesting problems of distributed control and optimization have appeared in several domains of applications, among which robotic networks, smart grids, transportation systems, multi agent systems etcetera. An example of a (linear) dynamic network is sketched in Figure \ref{fig:examnetw}, where excitation signals \red{$r$} and disturbance signals \red{$v$}, together with the linear dynamic modules \red{$G$} induce the behaviour of the node signals \red{$w$}.

When structured systems like the one in Figure \ref{fig:examnetw} become of interest for analysing performance and stability, it is appropriate to also consider the development of (data-driven) models.
In system identification literature, where the majority of the work is focused on open-loop or feedback controlled (multivariable) systems, there is an increasing interest in data-driven modeling problems related to dynamic networks. Particular questions that can be addressed are, e.g.:
\begin{itemize}
\item[(a)] Identification of a single selected module $G_{ji}$, on the basis of measured \red{signals} $w$ and $r$;
\item[(b)] Identification of the full network dynamics;
\item[(c)] Identification of the topology of the network, i.e. the Boolean interconnection structure between the several nodes $w_i$.
\end{itemize}

The problem (a) of identifying a single module in a dynamic network has been addressed in \citep{VandenHof&Dankers&etal:13}, where a framework has been introduced for prediction error identification in dynamic networks, and classical closed-loop identification techniques have been generalized to the situation of structured networks.
Using this framework, predictor input selection (\citep{Dankers&etal_TAC:16}) has been addressed to decide on which node signals need to be measured for identification of a particular network module. Errors-in-variables problems have been addressed in (\citep{Dankers&etal_Autom:15}) to deal with the situation when node signals are measured subject to additional sensor noise.

The problem (b) of identifying the full network can be recast into a multivariable identification problem, that can then be addressed with classical identification methods \cite{Soderstrom&Stoica:89}. Either structured model sets can then be used, based on an a priori known interconnection structure of the network, or a fully parametrized model set, accounting for each and every possible link between node signals.

The problem (c) of topology detection has been addressed in e.g. \citep{MaterassiSalapaka2012} where Wiener filters have been used to reconstruct the network topology. In \citep{ChiusoPAuto2012} a Bayesian viewpoint has been taken and regularization techniques have been applied to obtain sparse estimates. Topology detection in a large scale network has been addressed in \citep{Sanandaji2011,sanandaji2012review} using compressive sensing methods, and in a biological network in \citep{Yuan2011,Yuan2012} using also sparse estimation techniques. Causal inference has been addressed in \citep{Quinn&etal:11}.

Not only in problem (b) but also in problem (c), the starting point is most often to model all possible links between node signals, in other words to parametrize all possible modules  $G_{ji}$ in the network. However when identifying such a full network model, care has to be taken that different network models can indeed be distinguished on the basis of the data that is available for identification. In  \citep{Goncalves08,adebayo2012dynamical} specific local conditions have been formulated for injectivity of the mapping from the network transfer function (transfer from external signals $r$ to node signals $w$) to network models.
This is done
outside an identification context and without considering (non-measured) disturbance inputs.
Uniqueness properties of a model set for purely stochastic networks (without external excitations $r$) have been studied in \citep{MaterassiSalapaka2012,Hayden&etal:13} where the assumption has been made, like in many of the works in this domain, that each node is driven by an independent white noise source.

In this paper we are going to address the question: under which conditions on \red{the} experimental setup and choice of model set, different network models in the set can be distinguished from each other on the basis of measured data? The typical conditions will then include presence and location of external excitations, presence of and modelled correlations between disturbance signals, and modelled network topology. \\
This question will be addressed by introducing the concept of network identifiability as a property of a parametrized set of network models. We will study this question for the situations that
\begin{itemize}
\item Disturbance terms $v_i$ are allowed to be correlated over time but also over node signals, i.e. $v_i$ and $v_j$, $i \neq j$ can be correlated.
\item The vector disturbance process $v := [v_1^T\ v_2^T \cdots ]^T$ can be of reduced-rank, i.e. has a driving white noise process that has a dimension that is strictly less than the dimension of $v$. This includes the situation that some disturbance terms can be $0$.
\item Direct feedthrough terms are allowed in the network modules.
\end{itemize}
%
The presence of possible correlations between disturbances, limits the opportunities to break down the modelling of the network into several multi-input single-output MISO) problems, as e.g. done in \citep{VandenHof&Dankers&etal:13}. For capturing these correlations among disturbances all relevant signals will need to be modelled jointly in a multi-input multi-output (MIMO) approach.

If the size of a dynamic network increases, the assumption of having a full rank noise process becomes more and more unrealistic. Different node signals in the network are likely to experience noise disturbances that are highly correlated with and possibly dependent on other node signals in its direct neighbourhood.
One could think e.g. of a network of temperature measurements in a spatial area, where unmeasured external effects (e.g. wind) affect all measured nodes in a strongly related way.
In the identification literature little attention is paid to this situation.
In a slightly different setting, the classical closed-loop system (Figure \ref{fig1}) also has this property, by considering the input to the process $G$ to be disturbance-free, rendering the two-dimensional vector noise process of reduced-rank. Closed-loop identification methods typically work around this issue by either replacing the external excitation signal $r$ by a stochastic noise process, as e.g. in the joint-IO method (\citep{caines1975feedback}), or by only focussing on predicting the output signal and thus identifying the plant model (and not the controller), as e.g. in the direct method (\citep{ljung:99}). In econometrics dynamic factor models have been developed to deal with the situation of high dimensional data and rank-reduced noise (\citep{Deistler2010211,Deistler2015}).

The notion of identifiability is a classical notion in system identification, but the concept has been used in different settings. The classical definition as present in \citep{ljung1976,Soderstrom&Ljung&Gustavsson:76} is a consistency-oriented concept concerned with estimates converging to the true underlying system (system identifiability) or to the true underlying parameters (parameter identifiability).
In the current literature, identifiability has become a property of a parametrized model set, referring to a unique one-to-one relationship between parameters and predictor model, see e.g. \citep{ljung:99}. As a result a clear distinction has been made between aspects of data informativity and identifiability. For an interesting account of these concepts see also the more recent work \citep{BazanellaGevers2010}.
In the current literature the structure/topology of the considered systems has been fixed and restricted to the common open-loop or closed-loop cases. In our network situation we have to deal with additional structural properties in our models. These properties concern e.g. the choices where external excitation and disturbance signals are present, and how they are modeled, whether or not disturbances can be correlated, and whether modules in the network are known and fixed, or parametrized in the model set.
In this paper we will particularly address the structural properties of networks, and we will introduce the concept of network identifiability, as the ability to distinguish networks models in identification. Rather than focussing on the uniqueness of parameters, we will focus on uniqueness of network models.

We are going to employ the dynamic network framework as described in \citep{VandenHof&Dankers&etal:13}, and we will introduce and analyse the concept of \textit{network identifiability} of a parametrized model set. We will build upon the earlier introduction of the problem and preliminary results presented in \citep{weerts_etal_2015} and \citep{weerts_etal_ALCOSP2016}, but we will reformulate the starting points and definitions, as well as extend the results to more general situations in terms of correlated noise, reduced-rank noise, and absence of delays in network modules.

This paper will proceed by defining the network setup (Section \ref{sect:netw}), and subsequently formulating the models, model sets and identifiability concept (Section \ref{sect:pred}). In Section \ref{sect:ident1}
\red{conditions to ensure network identifiability are presented for various situations,}
after which some examples are provided in Section \ref{sec:exam}. In Section \ref{sect:ident2} results are provided that exploit the particular interconnection structure that is present in the model set, after which a discussion section follows and conclusions are formulated.

\section{Dynamic network setting} \label{sect:netw}
Following the basic setup of \citep{VandenHof&Dankers&etal:13}, a dynamic network is built up out of $L$ scalar \emph{internal variables} or \emph{nodes} $w_j$, $j
= 1, \ldots, L$, and $K$ \emph{external variables} $r_k$, $k=1,\ldots K$.
Each internal variable is described as:
\begin{align}
w_j(t) = \sum_{\stackrel{l=1}{l\neq j}}^L
G_{jl}(q)w_l(t) + \sum_{k=1}^K
R_{jk}(q)r_k(t) + v_j(t)
\label{eq:netw_def}
\end{align}
where $q^{-1}$ is the delay operator, i.e. $q^{-1}w_j(t) = w_j(t-1)$;
\begin{itemize}[leftmargin=*]
	\item $G_{jl}$, $R_{jk}$ are proper rational transfer functions, and the single transfers $G_{jl}$ are referred to as {\it modules} in the network.
	\item $r_k$ are \emph{external variables} that can directly be manipulated by the user;
	\item $v_j$ is \emph{process noise}, where the vector process $v=[v_1 \cdots v_L]^T$ is modelled as a stationary stochastic process with rational spectral density, such that there exists a $p$-dimensional white noise process $e:= [e_1 \cdots e_p]^T$, $p \leq L$, with covariance matrix $\Lambda>0$ such that
\[ v(t) = H(q)e(t). \]
\end{itemize}

The noise model $H$ requires some further specification. For $p=L$, referred to as the full-rank noise case, $H$ is square, stable, monic and minimum-phase. The situation $p < L$ will be referred to as the \emph{singular} or \emph{rank-reduced} noise case. In this latter situation it will be assumed that
the $L$ node signals $w_j$, $j=1, \cdots L$ are ordered in such a way that the first $p$ nodes are affected by a full-rank noise process, thus allowing a representation for $H$ that satisfies
\begin{equation}
\label{eqH}
H(q) = \begin{bmatrix} H_a \\ H_b \end{bmatrix}
\end{equation}
with $H_a$ square and  monic, while $H$ is stable and has a stable left inverse $H^\dagger$, satisfying $H^\dagger H = I_p$, the $p \times p$ identity matrix.

When combining the $L$ node signals we arrive at the full network expression
\begin{align*}
\begin{bmatrix}  \! w_1 \!  \\[1pt] \! w_2 \!  \\[1pt]  \! \vdots \! \\[1pt] \! w_L \!  \end{bmatrix} \!\!\! = \!\!\!
\begin{bmatrix}
0 &\! G_{12} \!& \! \cdots \! &\!\! G_{1L} \!\\
\! G_{21} \!& 0 & \! \ddots \! &\!\!  \vdots \!\\
\vdots &\! \ddots \!& \! \ddots \! &\!\! G_{L-1 \ L} \!\\
\! G_{L1} \!&\! \cdots \!& \!\! G_{L \ L-1} \!\! &\!\! 0
\end{bmatrix} \!\!\!\!
\begin{bmatrix} \! w_1 \!\\[1pt]  \! w_2 \!\\[1pt] \! \vdots \!\\[1pt] \! w_L \! \end{bmatrix} \!\!\!
+ \!\! R \! (q) \!\!\!
\begin{bmatrix} \! r_1 \!\\[1pt] \! r_2 \!\\[1pt] \! \vdots \!\\[1pt]  \! r_{K} \!\end{bmatrix}
\!\!\!+\!\!
H \! (q) \!\!\! \begin{bmatrix}\! e_1 \!\\[1pt] \! e_2 \!\\[1pt] \! \vdots \!\\[1pt] \! e_p\!\end{bmatrix} \!\!\!
\end{align*}
Using obvious notation this results in the matrix equation:
\begin{align} \label{eq.dgsMatrix}
w = G w + R r + H e.
\end{align}
The network transfer function that maps the external signals $r$ and $e$ into the node signals $w$ is denoted by:
\begin{equation}
\label{eq:T0}
T(q) = \begin{bmatrix} T_{wr}(q) & T_{we}(q) \end{bmatrix},
\end{equation}
with
\begin{align}
&T_{wr}(q) := \left( I-G(q)\right)^{-1}R(q), \ \mbox{and}& \label{eq:twr}\\
&T_{we}(q) := \left( I-G(q)\right)^{-1}H(q).& \label{eq:twe}
\end{align}
%
%
The identification problem to be considered is the problem of identifying the network dynamics ($G, R, H, \Lambda$) on the basis of \red{measured variables $w$ and $r$}.

The dynamic network formulation above is related to what has been called the {\it Dynamic Structure Function (DSF)} as considered for disturbance-free systems in \citep{adebayo2012dynamical,Yuan2011,Yuan2012}. In particular, state space structures can be included by considering every module to be restricted to having first order dynamics only.\\
\red{In terms of notation, for any transfer function $A(z)$ we will denote $A^\infty := \lim_{z\rightarrow\infty} A(z)$}.

\section{Network model set and identifiability} \label{sect:pred}
In order to arrive at a definition of network identifiability we need to specify a network model and a network model set.
\begin{definition}[network model]
\label{def1}
A network model of a network with $L$ nodes, and $K$ external excitation signals, with a noise process of rank $p \leq L$ is defined by the quadruple:
\[ M = (G,R,H,\Lambda) \]
with
\begin{itemize}
\item $G \in \R^{L \times L}(z)$, diagonal entries 0, all modules proper and stable\footnote{The assumption of having all modules stable is made in order to guarantee that $T_{we}$ (\ref{eq:twe}) is a stable spectral factor of the noise process that affects the node variables.};
\item $R \in \R^{L \times K}(z)$, proper;
\item $H \in \R^{L \times p}(z)$, stable, with a left stable inverse, and satisfying (\ref{eqH}).
\item    $\Lambda \in \R^{p\times p}$, $\Lambda > 0$;
\item the network is well-posed\footnote{This implies that all \red{principal} minors of $(I-G(\infty))^{-1}$ are nonzero.}  \citep{Dankers_diss}, with $(I-G)^{-1}$ proper and stable. \hfill $\Box$
\end{itemize}
\end{definition}
We include the noise covariance matrix $\Lambda$ in the definition of a model, as is common for multivariable models \citep{Soderstrom&Stoica:89}. The noise model $H$ is defined to be non-square in the case of a rank-reduced noise ($p<L$).
\begin{definition}[network model set]
\label{def2}
A network model set for a network of $L$ nodes, $K$ external excitation signals, and a noise process of rank $p \leq L$, is defined as a set of parametrized matrix-valued functions:
\[ \M := \left\{ M(\theta) = \bigl(G(q,\theta), R(q,\theta), H(q,\theta), \Lambda(\theta)\bigr), \theta \in \Theta \right\}, \]
with all models $M(\theta)$ satisfying the properties as listed in Definition \ref{def1}.\hfill $\Box$
\end{definition}
In this paper we will consider model sets for which all models in the set share the same rank, i.e. $rank(\Lambda(\theta))=p$. We will use parameters $\theta$ only as a vehicle for creating a set of models. We will not consider any particular properties of the mapping from parameters to network models.

\red{The question} whether in a chosen model set, the models can be distinguished from each other through identification, has two important aspects:
\begin{itemize}
\item a structural ---or identifiability--- aspect: is it possible at all to distinguish between models, given the presence and location of external excitation signals and noise disturbances, and
\item a data informativity aspect: given the presence and location of external excitation signals and noise disturbances, are the actual signals informative enough to distinguish between models during a particular identification experiment.
\end{itemize}
We will refer to the first (structural) aspect as the notion of network identifiability. For consistency of model estimates in an actual identification experiment, it is then required that the model set is network identifiable and that the external excitation signals are sufficiently informative. This separation of concepts allows us to study the structural aspects of networks, separate from the particular choice of test signals in identification.

Based on the network equations (\ref{eq.dgsMatrix})-(\ref{eq:twe}) we can rewrite the system as
\begin{eqnarray}
w & = & T_{wr}(q)r(t) + \bar v(t) \label{eq:n1} \\
\mbox{where }\ \ \
\bar v(t) & := & T_{we}(q)e(t).\label{eq:n2}
\end{eqnarray}
Many identification methods, among which prediction error and subspace identification methods, base their model estimates on second order statistical properties of the measured data. These properties are represented by auto-/cross-correlation functions or spectral densities of the signals $w$ and $r$. On the basis of the expressions (\ref{eq:n1})-(\ref{eq:n2}), and noticing that $r$ is measured and $e$ is not, the model objects that generate the second-order properties of $w$, are typically given by the transfer function $T_{wr}(q)$ and the spectral density $\Phi_{\bar v}(\omega)$, with $\Phi_{\bar v}(\omega)$, being defined as
$\Phi_{\bar v}(\omega) := \mathcal{F}\{\mathbb{E} [\bar v(t)\bar v^T(t-\tau)]\}$, where $\mathcal{F}$ is the discrete-time Fourier transform, and $\mathbb{E}$ the expected value operator.
%
By utilizing (\ref{eq:twr})-(\ref{eq:twe}), we can now write for a parametrized model $M(\theta)$:
\begin{eqnarray*}
T_{wr}(q,\theta) & := & [I-G(q,\theta)]^{-1}R(q,\theta), \\
\Phi_{\bar v}(\omega,\theta) & = & [I-G(\eio,\theta)]^{-1}H(\eio,\theta)\Lambda(\theta)\cdot \\ & & \hspace*{1cm} \cdot H(\eio,\theta)^*[I-G(\eio,\theta)]^{-*},
\end{eqnarray*}
where $(\cdot)^*$ denotes complex conjugate transpose. As a result we arrive at a definition of network identifiability that addresses the property that network models are uniquely determined from $T_{wr}$ and $\Phi_{\bar v}$.
\begin{definition}[Network identifiability]
\label{defif}
The network model set $\M$ is globally network identifiable at $M_0 :=M(\theta_0)$ if for all models $M(\theta_1) \in \M$,
\begin{equation} \label{equivTP}
		\left. \begin{array}{c} T_{wr}(q,\theta_1) = T_{wr}(q,\theta_0) \\ \Phi_{\bar v}(\omega,\theta_1) = \Phi_{\bar v}(\omega,\theta_0) \end{array} \right\}
		\Rightarrow
		M(\theta_1) = M(\theta_0).
\end{equation}
$\M$ is globally network identifiable if (\ref{equivTP}) holds for all $M_0 \in \M$.\hfill $\Box$
\end{definition}
%
We have chosen to use the spectral density $\Phi_{\bar v}$ in the definition, rather than its spectral factor as e.g. done in \cite{weerts_etal_2015}. This is motivated by the objective to include the situation of rank-reduced noise, where $\Phi_{\bar v}(\omega,\theta)$ will be singular, and the handling of possible direct feedthrough terms and algebraic loops in the network. This will be further addressed and clarified in Section \ref{sect:ident1}.
\begin{remark} \label{remark1}
In the definition we consider identifiability of the network dynamics $M=(G,R,H,\Lambda)$. This can simply be generalized to consider the identifiability of a particular network property $f(M)$, by replacing the right hand side of the implication (\ref{equivTP}) by
$f(M_1) = f(M_0)$, while $f$ can refer to network properties as e.g. the Boolean topology of the network, or the network dynamics in $G$, etcetera. \hfill $\Box$
\end{remark}

\begin{remark}
\red{The definition allows the handling of several situations, using either signals $w$ as data, or the combined signals $w$ and $r$.
Note that e.g. in} the direct method and joint-io method for closed-loop identification \citep{ljung:99}, only the measured signals in $w$ are used as a basis for identifiability studies. In these approaches, excitation signals $r$ may be present, but are not taken into account. \red{ This situation can be handled by removing matrix $R$ from the model set.}
\hfill $\Box$
\end{remark}

Before moving to the formulation of verifiable conditions for network identifiability, we present an example of a disturbance free network to illustrate that a model set can be globally identifiable at one model, but not at another model. The example is taken from \citep{weerts_etal_2015}.
\begin{example} \label{example1}
	Consider the disturbance-free systems $\mathcal{S}_1$, $\mathcal{S}_2$ in Figure \ref{fig:example2sys} with $A(q) \! \neq \! 0,-1 $, and $B(q) \! \neq \! 0$, both rational transfer functions. These two networks are described by the transfer functions
\begin{equation}
		G_1^0 =
		\begin{bmatrix}
			0 & 0 & 0  \\
			A & 0 & 0  \\
			0 & B & 0
		\end{bmatrix},	
		\
    G_2^0 =
		\begin{bmatrix}
			0 & 0 & 0 \\
			0 & 0 & B \\
			A & 0 & 0
		\end{bmatrix},\
R_1^0=R_2^0 = \begin{bmatrix}
			1 & 0 \\
			0 & 1 \\
			1 & 0
		\end{bmatrix}. \nonumber
\end{equation}
The transfer function matrices $T_{wr}(q)$ related to the networks $\mathcal{S}_1$ and $\mathcal{S}_2$ respectively, are given by:
	\begin{equation}
 \label{eq:t12}
		T_1^0(q) \! = \!
		\begin{bmatrix}
			1 & 0 \\
			A & 1 \\
			AB+1 \; & B
		\end{bmatrix},\ \
		T_2^0(q) \! = \!
		\begin{bmatrix}
			1 & 0 \\
			 (A+1)B \; & 1 \\
			A+1 & 0
		\end{bmatrix}.
	\end{equation}
These transfer functions map the external signals $r$ to the node signals $w$.
We consider the model set $\mathcal{M}(\theta)$ with (omitting arguments $q$)
	\begin{equation}
     \label{eq:gr}
		G(\theta) =
		\begin{bmatrix}
			0 & G_{12}(\theta)  & G_{13}(\theta)  \\
			G_{21}(\theta)  & 0 & G_{23}(\theta)  \\
			G_{31}(\theta)  & G_{32}(\theta)  & 0
		\end{bmatrix},	
		 \ \ R =
		\begin{bmatrix}
			1 & 0 \\
			0 & 1 \\
			1 & 0
		\end{bmatrix},
	\end{equation}
and so $G(\theta)$ is parametrized and $R$ is \red{known and} fixed. Since we have a disturbance free system we discard a noise model here, without loss of generality.\\
In order to investigate whether each of the two systems can be represented uniquely within the model set, we refer to (\ref{eq:twr}), and analyze whether the equation
	\begin{equation}
		T_i^0(q) = [I-G(q,\theta)]^{-1} R(q) \label{eq:ex1mod}
	\end{equation}
for $i=1,2$ has a unique solution for $G(q,\theta)$. To this end we premultiply
(\ref{eq:ex1mod}) with $[I-G(q,\theta)]$. \\
For network $\mathcal{S}_1$ we then obtain the relation (omitting argument $q$)
	\begin{align}
		&\begin{bmatrix}
			1 & \text{-}G_{12}(\theta) & \text{-}G_{13}(\theta) \\
			\text{-}G_{21}(\theta) & 1 & \text{-}G_{23}(\theta) \\
			\text{-}G_{31}(\theta) &  \text{-}G_{32}(\theta) & 1
		\end{bmatrix}	\!\!
		\begin{bmatrix}
			1 & 0 \\
			A & 1 \\
			AB+1 & B
		\end{bmatrix}
		=
		\begin{bmatrix}
			1 & 0 \\
			0 & 1 \\
			1 & 0
		\end{bmatrix}	.
		\label{eq:ex_mod_sys1}
	\end{align}
Solving the corresponding six equations for the parametrized transfer functions $G_{ij}(\theta)$ shows the following. When combining the two equations related to the first row in the right hand side matrix of (\ref{eq:ex_mod_sys1}) it follows that $G_{13}(\theta)=G_{12}(\theta)=0$. Solving the second row leads to $G_{23}(\theta)=0$ and $G_{21}(\theta)=A$, while solving the third row delivers
$G_{31}(\theta)=0$ and $G_{32}(\theta)=B$. As a result
%
the original system $\mathcal{S}_1$ is uniquely recovered, and so $\M$ is globally network identifiable at $\mathcal{S}_1$.\\
%
When applying the same reasoning to network $\mathcal S_2$ we obtain
	\begin{align}
		&\begin{bmatrix}
			1 & \text{-}G_{12}(\theta) & \text{-}G_{13}(\theta) \\
			\text{-}G_{21}(\theta) & 1 & \text{-}G_{23}(\theta) \\
			\text{-}G_{31}(\theta) &  \text{-}G_{32}(\theta) & 1
		\end{bmatrix}	\!\!
		\begin{bmatrix}
			1 & 0 \\
			(A+1)B & 1 \\
			A+1 & 0
		\end{bmatrix}
		=
		\begin{bmatrix}
			1 & 0 \\
			0 & 1 \\
			1 & 0
		\end{bmatrix}.
		\label{eq:ex_mod_sys2}
	\end{align}
Solving this system of equations for the second column on the right hand side leads to $G_{12}(\theta)=G_{32}(\theta)=0$, while the solution for the first column delivers
$G_{13}(\theta)=0$, $G_{31}(\theta)=A$ and
	\begin{align}
	&-G_{21}(\theta) +(A+1)B - G_{23}(\theta)(A+1) = 0 \label{eq:ex_uonunique}
	\end{align}
or equivalently
$ G_{21}(\theta) = (A+1)(B-G_{23}(\theta)). $
This shows that not only $G_{21}(\theta)=0$, $G_{23}(\theta)=B$ is a valid solution, but actually an infinite number of solutions exists. As a result $\M$ is not globally network identifiable at $\mathcal{S}_2$.
	An interpretation is that in $\mathcal{S}_2$ the contributions from $w_1$ and $w_3$ both solely depend on $r_1$ making them indistinguishable, which is reflected in the modeled transfer function matrix $R(q)$. \hfill $\Box$
	\begin{figure}[t]
		\centering
        \includegraphics[height=0.45\columnwidth]{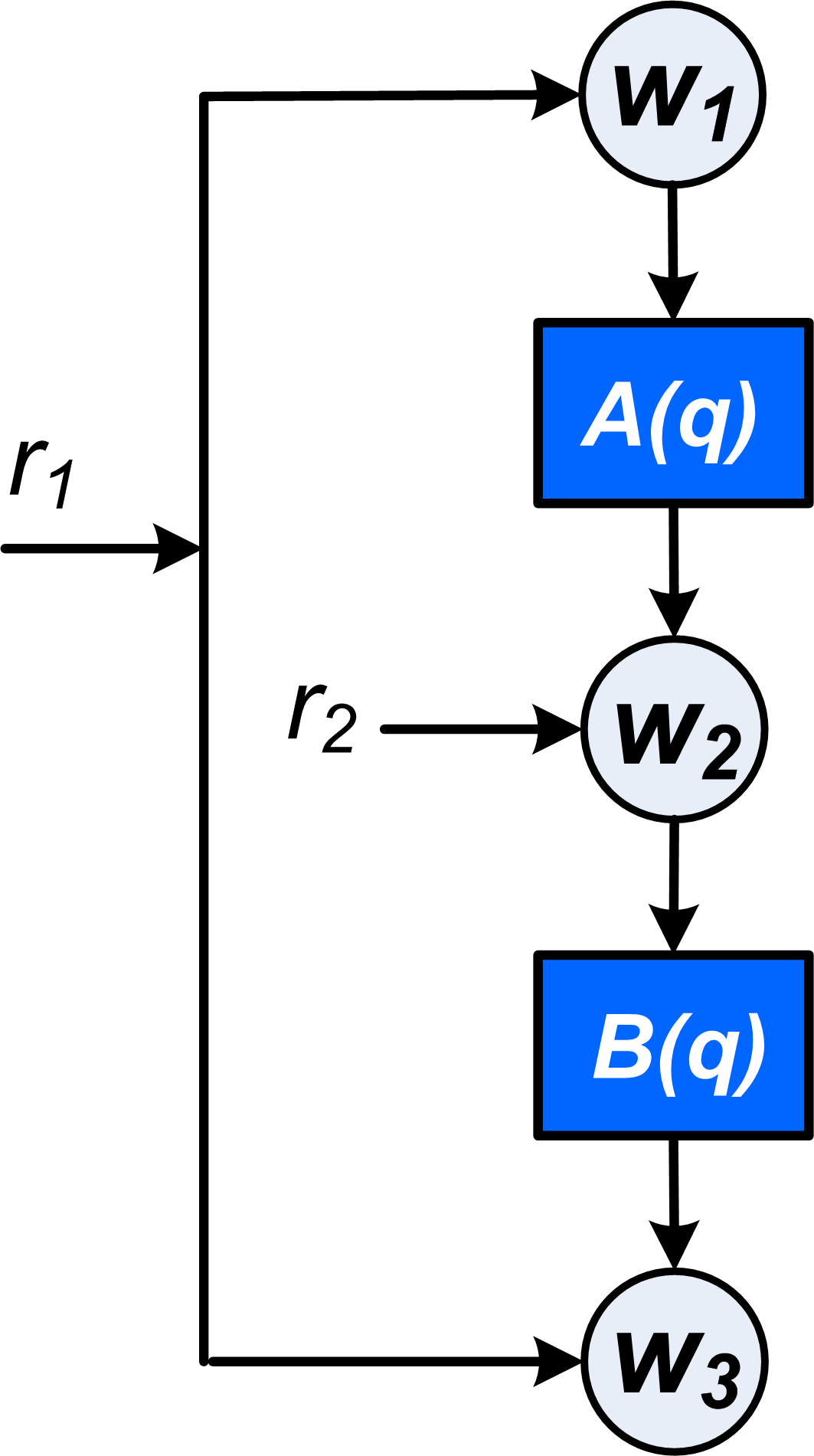}\hspace*{1cm}
        \includegraphics[height=0.45\columnwidth]{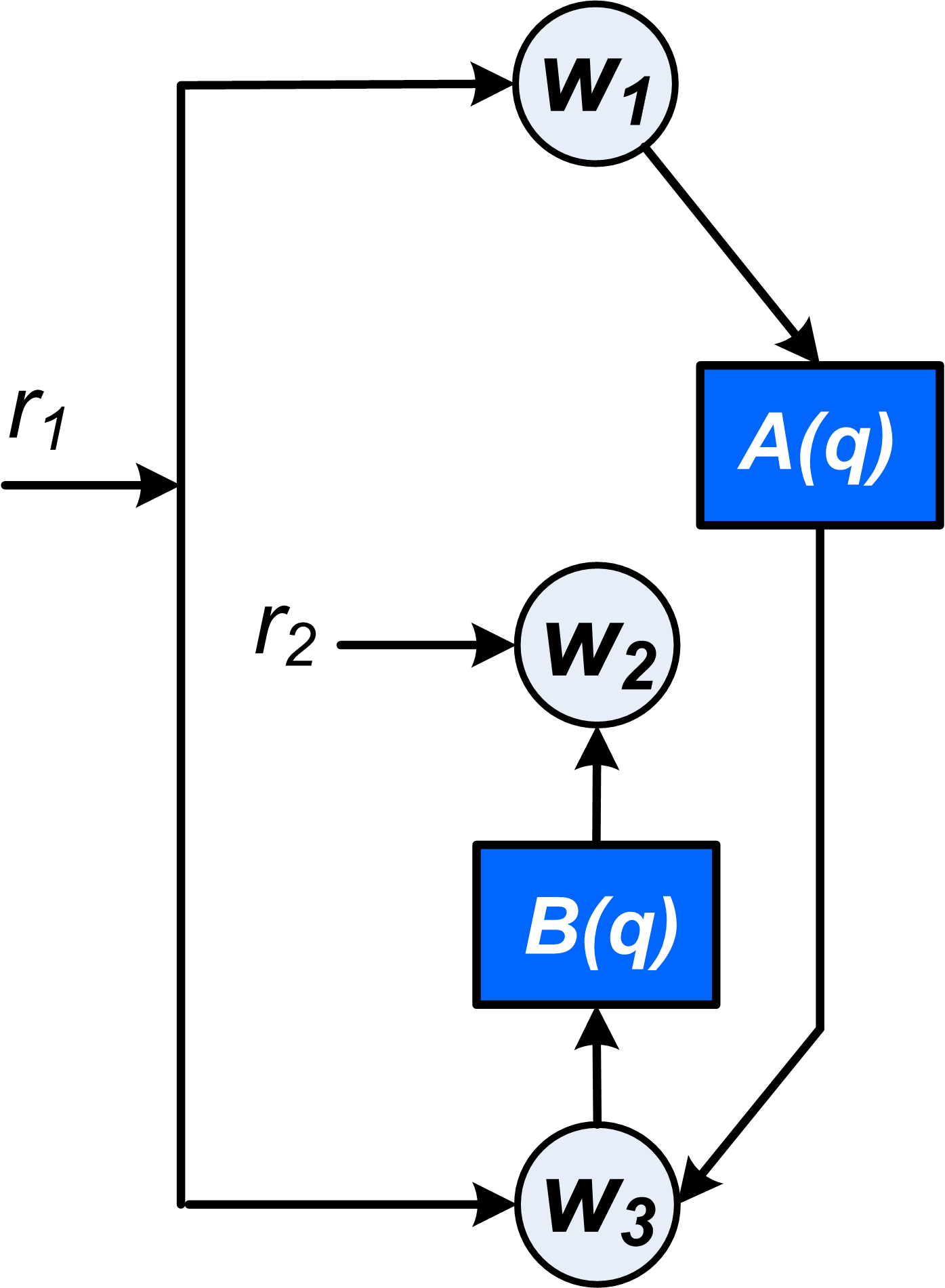}
		\caption{Systems $\mathcal{S}_1$ (left) and $\mathcal{S}_2$ (right).}
		\label{fig:example2sys}
	\end{figure}
\end{example}
\begin{remark}
In the setting of an identification problem of either the dynamics or the topology of the network, it will be most important to be able to verify whether {\it all} models in a particular model set can be distinguished from each other, rather than the \red{identifiability} of one particular model in the set. Since in an identification problem the underlying real data generating network is unknown, global network identifiability of a model set $\M$ \red{is a stronger concept} than global network identifiability at a particular model. The latter property is \red{considered} in \citep{Goncalves08} and \citep{Gevers:16}. \red{Here we will address both properties.} \hfill $\Box$
\end{remark}

\section{Conditions for verifying network identifiability}   \label{sect:ident1}

In this Section we will derive conditions for verifying global network identifiability.
To this end the implication (\ref{equivTP}) of Definition \ref{defif} will be reformulated into a condition on the network transfer functions $T(q,\theta)$, that is more easy to verify. This reformulation is done for three different situations, specifying particular assumptions on the presence/absence of delays in the modules in the networks.

First we \red{are making} the following assumption:
\begin{assumption} \label{asspi}
We will consider model sets that satisfy the property that all models in the set share the same value of rank $\Lambda(\theta) = p$, and the node signals $w$ are ordered in such a way that for all models in the set the first $p$ components of $v$ constitute a full rank process.
\end{assumption}

This Assumption may look rather restrictive, but actually it can be shown that for the analysis of network identifiability {\it at a particular model}, the assumption is not restrictive. Additionally the required value of $p$ as well as the requested ordering of signals can be determined from data. This topic will be further addressed in Section \ref{sec:disc}.

Before being able to formulate verifiable conditions for identifiability, we need to collect some properties of reduced-rank spectra, in order to properly handle the situation that $p < L$.

\subsection{Factorizations of reduced-rank spectra}

\begin{lemma}[reduced-rank spectra]
\label{lemrr}
Consider an $L$-dimensional stationary stochastic process $x$ with rational spectral density $\Phi_x$ and rank $p < L$, that satisfies the ordering property of Assumption \ref{asspi}. Then
\begin{itemize}
\item[a.] $\Phi_x$ allows a unique spectral factorization
\[ \Phi_x = F \Delta F^* \]
with $F \in \R^{L\times p}(z)$, $F = \begin{bmatrix} F_a \\ F_b \end{bmatrix}$ with $F_a$ square, monic, and $F$ stable and having a stable left inverse $F^\dagger$ that satisfies $F^\dagger F = I_p$, $\Delta \in \R^{p\times p}$, $\Delta > 0$;
\item[b.] Based on the unique decomposition of $\Phi_x$ in (a.), there exists a unique factorization of $\Phi_x$ in the structure:
\[ \Phi_x = \breve F \breve\Delta \breve F^* \]
with $\breve F$ square and monic, and $\breve \Delta \in \R^{L \times L}$, having the particular structure
\[
\breve F = \begin{bmatrix} F_a & 0 \\ F_b-\Gamma & I \end{bmatrix},\ \ \ \ \
 \breve \Delta = \begin{bmatrix} I \\ \Gamma \end{bmatrix} \Delta \begin{bmatrix} I \\ \Gamma \end{bmatrix}^T
\]
and $\Gamma := \lim_{z\rightarrow\infty} F_b(z)$;
\end{itemize}
\end{lemma}

{\bf Proof.} Part (a) is the standard spectral factorization theorem, see \cite{Youla:61}. Part (b) can be verified by direct computation.

\red{This spectral factorization result shows that for the modelling of the noise process $v$, we actually have two options. The first is a noise model $v = H e$ with $e$ a $p$-dimensional (full-rank) noise process, and a $L \times p$ noise filter $H$ structured as $H = \begin{bmatrix} H_a \\ H_b \end{bmatrix}$ of which the upper square part $H_a$ is monic. The second option is a noise model $v = \breve H \breve e$, with $\breve e$ an $L$-dimensional (possibly reduced rank) noise process, and $\breve H$ a monic square noise filter, structured as $\begin{bmatrix} H_a & 0 \\ H_b-H_b^\infty & I\end{bmatrix}$. In this paper we will dominantly use the first (non-square) representation, while the second (square) representation will be effectively utilized in many of the proofs.
}

Now we are up to formulating conditions for network identifiability. To this end we will distinguish between different situations, dependent on the presence of delays in the network.

\subsection{The situation of strictly proper modules}

First we consider the situation that all modules in the network are strictly proper, i.e. $\lim_{z\rightarrow\infty} G(z) = 0$.

\begin{proposition}
\label{propT}
Consider a network model set $\M$, and define
\begin{equation}
\label{eq:t}
T(q,\theta):= (I-G(q,\theta) )^{-1} U(q,\theta)
\end{equation}
\begin{equation}
\label{eq:u}
\mbox{with } U(q,\theta):= \begin{bmatrix} R(q,\theta) & H(q,\theta) \end{bmatrix}
\end{equation}
with $T(q,\theta)$ being the parameterized version of the network transfer function $T(q)$ (\ref{eq:T0}).
If
\begin{itemize}
\item $G^\infty(\theta) := \lim_{z\rightarrow\infty} G(z,\theta)=0$ for all $\theta \in \Theta$,
\end{itemize}
then condition (\ref{equivTP}) in Definition \ref{defif} of network identifiability is equivalently formulated as
	\begin{eqnarray} \label{equivT}
		\lefteqn{\{T(q,\theta_1) = T(q,\theta_0)\}
		\Rightarrow} \\
		& & \ \ \ \{(G(\theta_1), R(\theta_1), H(\theta_1)) = (G(\theta_0),R(\theta_0), H(\theta_0))\}. \nonumber
	\end{eqnarray}
\end{proposition}
The proof is provided in the Appendix.

Note that the above result is valid for both full-rank ($p=L$) and reduced-rank ($p<L$) noise processes. Additionally there are no restrictions on $\Lambda(\theta)$, e.g. it is not restricted to being diagonal. In \cite{Goncalves08} the transfer function $T$ has been used as a basis for dynamic structure reconstruction, \red{in a continuous-time domain setting.}
The fact that the network transfer function $T$ is the object that can be uniquely identified from data,
\red{has been analyzed in \cite{weerts_etal_2015} for the situation that $p=L$ with diagonal $\Lambda(\theta)$, and no algebraic loops in the networks.} This has been the motivation in \cite{weerts_etal_2015} to use the condition (\ref{equivT}) as a {\it definition} of network identifiability. In the situation of rank-reduced noise, including noise-free nodes, this result is still true under the formulated condition that all modules in the network are strictly proper. The only adaptation is that the transfer function $T_{we}(q,\theta)$ is no longer square but rectangular in its dimension, i.e. $L\times p$.
An equivalent formulation of (\ref{equivT}) is obtained by adding the equality of covariance matrices $\Lambda$ to both sides of the implication, leading to
\begin{equation} \label{equivTl}
	\left. \begin{array}{c} T(q,\theta_1) = T(q,\theta_0) \\ \Lambda(\theta_1) = \Lambda(\theta_0) \end{array}
		\right\} \Rightarrow M(\theta_1) = M(\theta_0).
	\end{equation}
In this representation it is clear that, when starting from expression (\ref{equivTP}) in the definition of network identifiability, $T_{we}(q,\theta)$ and $\Lambda(\theta)$ are uniquely determined from $\Phi_{\bar v}(\omega,\theta)$.
\subsection{The situation of modules with direct feedthrough}
In order to handle the situation of having direct feedthrough terms in $G$, we need to deal with the phenomenon of algebraic loops.

\begin{definition}
In a dynamic network there is an {\it algebraic loop} around node $w_{n_1}$, if there exists a sequence of integers $n_1,\cdots,n_k$ such that:
  \[ G_{n_1 n_2}^{\infty}G_{n_2 n_3}^{\infty} \cdots G_{n_k n_1}^{\infty} \neq 0, \] with $G_{n_1 n_2}^{\infty} := \lim_{z \rightarrow \infty}G_{n_1 n_2}(z)$.
\end{definition}

It can be shown (see \cite{Dankers_diss}) that there are no algebraic loops in a network if and only if there exists a permutation matrix $\Pi$, such that $\Pi^TG^\infty\Pi$ is upper triangular. We can now formulate a Proposition that is an alternative to Proposition \ref{propT}.

\begin{proposition}
\label{propT2}
Consider a network model set $\M$, and $T(q,\theta), U(q,\theta)$ according to (\ref{eq:t}),(\ref{eq:u}).
If
\begin{itemize}
\item there is no algebraic loop around any node signal in the parametrized model set, i.e. there exists a permutation matrix $\Pi$ such that for all $\theta\in \Theta$,
    \begin{equation} \Pi^TG^\infty(\theta)\Pi \end{equation}
    is upper triangular, and
\item $\Phi_v^\infty(\theta):= H^\infty(\theta)\Lambda(\theta)H^\infty(\theta)^T$ is diagonal for all $\theta\in\Theta$,
\end{itemize}
then condition (\ref{equivTP}) in Definition \ref{defif} of network identifiability is equivalently formulated as
(\ref{equivTl}).
\end{proposition}
The proof is provided in the Appendix.

For the particular situation of noise-free nodes, the result of this proposition has been applied in \cite{weerts_etal_ALCOSP2016}. Proposition \ref{propT2} in relation to Proposition \ref{propT}, shows that the ability to estimate more flexible correlations between the white noise processes ($\Phi_v^\infty(\theta)$ \red{is} not constrained in Proposition \ref{propT}, while being diagonal in Proposition \ref{propT2}), is traded against the ability to handle direct feedthrough terms in the modules (Proposition \ref{propT2}). It also should be noted that the above results hold true for any particular experimental setup, i.e. for any selection of excitation signals $r$ that are present.

\subsection{The situation of algebraic loops}
\label{sec:alglp}

The results of Proposition \ref{propT} and \ref{propT2} have been derived based on conditions that guarantee that the transfer function $T_{we}$ uniquely determines the model terms $(H,\Lambda)$. So actually this has been a reasoning that is fully based on the noise spectrum $\Phi_{\bar v}$. By incorporating more specific conditions on $T_{wr}$, more generalized situations can be handled, even including the situation of having algebraic loops in the network. We will follow a reasoning where the transfer function $T_{wr}$ will be required to uniquely determine the feedthrough term $G^{\infty}$, and --as a result--- also the noise covariance matrix $\Lambda$.

\red{To this end we consider the direct feedthrough terms $T_{wr}^\infty(\theta)$, $R^\infty(\theta)$ and $G^\infty(\theta)$.}
Suppose that row $i$ of $(I-G^\infty(\theta))$ has $\alpha_i$ parameterized elements, and row $i$ of $R^\infty(\theta)$ has $\beta_i$ parameterized elements.
We define the $L \times L$ permutation matrix $P_i$ and the $K \times K$ permutation matrix $Q_i$ such that all parametrized entries in the considered row  of $(I-G^\infty(\theta))\red{P_i}$ are \red{gathered on the left hand side}, and all parametrized entries in the considered row of $R^\infty(\theta)\red{Q_i}$ are \red{gathered on the right hand side}, i.e.
\begin{align}
	(I-G^\infty(\theta))_{i\star} P_i &= \begin{bmatrix} (I-G^\infty(\theta))^{(1)}_{i\star} & (I-G^\infty)^{(2)}_{i\star}	 \end{bmatrix} \label{eq:pi}
	\\
	R^\infty(\theta)_{i\star} Q_i &= \begin{bmatrix} {R^\infty}^{(1)}_{i\star} & {R^\infty}^{(2)}_{i\star}(\theta)	 \end{bmatrix} \label{eq:qi}
\end{align}	
with $(\cdot)_{i\star}$ indicating the $i$-th row of a matrix.

Next we define the matrix $\check T_i^\infty(\theta)$ of dimension $\alpha_i \times (K - \beta_i)$ as the submatrix of $T_{wr}^\infty(\theta)$ that is constructed by taking the row numbers that correspond to the columns of $G^\infty(\theta)_{i\star}$ that are parametrized, and by taking the column numbers that correspond to the columns of $R^\infty(\theta)$ that are not parametrized. This is formalized by
\begin{equation} \label{eq:tinf}
	\check T_i^\infty(\theta) :=
	\begin{bmatrix} I_{\alpha_i} & 0	\end{bmatrix} P_i^{-1} T_{wr}^\infty(\theta) Q_i \begin{bmatrix} I_{K-\beta_i} \\ 0	\end{bmatrix}.
\end{equation}
We can now formulate the following identifiability result for the situation that even algebraic loops are allowed in the network.
\begin{proposition}
\label{propT3}
Consider a network model set $\M$, and $T(q,\theta), U(q,\theta)$ according to (\ref{eq:t}),(\ref{eq:u}).
If for all $\theta\in\Theta$:
\begin{itemize}
\item each row of $\begin{bmatrix} G^\infty(\theta) & R^\infty(\theta)	\end{bmatrix}$ has at most $K$ parameterized elements, and
\item for each $i=1,\cdots L$, the matrix $\check T_i^\infty(\theta)$ has full row rank,
\end{itemize}
then condition (\ref{equivTP}) in Definition \ref{defif} of network identifiability is equivalently formulated as
(\ref{equivTl}).
\end{proposition}
The proof is provided in the Appendix.

In the Proposition, conditions are formulated under which the transfer function $T_{wr}$ will uniquely determine the direct-feedthrough term $G^\infty$ and ---as a result thereof--- also the noise covariance matrix $\Lambda$. \red{In a context of consistent identification methods, handling the situation} of algebraic loops is further discussed in \cite{Weerts&etal_CDC:16}.

\subsection{Network identifiability results for full excitation}

We have shown under which conditions the essential condition for global network identifiability can be equivalently formulated in the expression (\ref{equivTl}) on the basis of $T$ and $\Lambda$. We continue with showing when the implication (\ref{equivTl}) is satisfied in the situation that we have at least as many external excitation plus white noise inputs, as we have node signals. This leads to sufficient conditions for global network identifiability that are not dependent on the particular structure of the network as present in $G$.
\begin{theorem}\label{theo1a}
Let $\M$ be a network model set for which the conditions of one of the Propositions \ref{propT}-\ref{propT3} are satisfied. Then
\begin{itemize}
\item[(a)] $\M$ is globally network identifiable at $M(\theta_0)$ if there exists a nonsingular and parameter-independent transfer function matrix $Q\!  \in \!  \mathbb R^{(K + p)   \times  (K + p)}(z)$ such that
	\begin{equation}\label{eqt2} U(q,\theta)
Q(q) = \begin{bmatrix} D(q,\theta) & F(q,\theta) \end{bmatrix} \end{equation}
with $D(\theta)\in\mathbb R^{L\times L}(z)$, $F(\theta)\in\mathbb R^{L\times (p+K-L)}(z)$, and $D$ diagonal and full rank for all $\theta\in \Theta_0$ with
\[ \Theta_0 := \{\theta\in \Theta\ |\ T(q,\theta) = T(q,\theta_0)\}. \]
\item[(b)] If in part (a) the diagonal and full rank property of $D(q,\theta)$ is extended to all $\theta\in \Theta$, then $\M$ satisfies the condition (\ref{equivTl}) for global network identifiability at all $M(\theta_0)\in \M$.
\end{itemize}
\end{theorem}
The proof of the Theorem is added in the appendix.\\
Expression (\ref{eqt2}) is
basically equivalent to a related result in \citep{Goncalves08}, where a deterministic reconstruction problem is considered on the basis of a network transfer function, however without considering (non-measured) stochastic disturbance signals.
Note that the condition can be interpreted as the possibility to give $U(q,\theta)$ a leading diagonal matrix by column operations.
There is an implicit requirement in the theorem that $U$ has full row rank, and therefore it does not apply to the case of Example \ref{example1}. The situation of a rank-reduced matrix $U$ will be considered in Section \ref{sect:ident2}.
\begin{example} \label{example_corr_noise}
	Suppose we model correlated noise by having off-diagonal terms in $H$, in the model set  $\mathcal M(\theta)$ with
\begin{eqnarray*}
  &G\! =\! \begin{bmatrix} 0&G_{12}(\theta)&G_{13}(\theta)\\G_{21}(\theta)&0&G_{23}(\theta)\\G_{31}(\theta)&G_{32}(\theta)&0\end{bmatrix}\! , \;& \\ & H\! =\! \begin{bmatrix} H_{11}(\theta)&H_{12}(\theta)&0\\H_{21}(\theta)&H_{22}(\theta)&0\\0&0&H_{33}(\theta)\end{bmatrix}\! ,\; R\! =\! \begin{bmatrix} R_{11}(\theta)&0\\0&R_{22}(\theta)\\0&0\end{bmatrix}\! ,
\end{eqnarray*}
	where $R_{11}(\theta), R_{22}(\theta) \not\equiv 0$, and $H(\theta)$ monic.
	Then a simple permutation matrix $Q$ can be found to create $U(q,\theta)Q= \begin{bmatrix}	 D(q,\theta) & F(q,\theta)	 \end{bmatrix}$ with
	$ D(q,\theta) = \mathrm{diag}(\begin{bmatrix}R_{11}(\theta) & R_{22}(\theta)  & H_{33}(\theta)	 \end{bmatrix}) $	
	and by Theorem \ref{theo1a} the model set is globally network identifiable. If external excitation signals $r$ would have been absent, identifiability can not be \red{guaranteed according to Theorem \ref{theo1a}} because the off-diagonal terms in the noise model would prevent the existence of a permutation matrix $Q$ that can turn the noise model into a diagonal form. If the process noises at the first two nodes are uncorrelated, they can be modelled with $H_{21}(\theta)\equiv H_{12}(\theta)\equiv 0$, and the diagonal $H$ directly implies global network identifiability, irrespective of the presence of external excitation signals.
\hfill $\Box$
\end{example}
One of the important consequences of Theorem \ref{theo1a} is formulated in the next corollary.
\begin{corollary}
Subject to the conditions in Theorem \ref{theo1a}, a network model set $\M$ is globally network identifiable if every node signal in the network is excited by either an external excitation signal $r$ or a noise signal $v$, that is uncorrelated with the excitaton/noise signals on the other nodes.
\end{corollary}
The situation described in the Corollary corresponds to $U(q,\theta)$ having a single parametrized entry in every row and every column, and thus implies that $U(q,\theta)$ can be permuted to a diagonal matrix.
Uncorrelated excitation can come from noise or external variables.
Note that the result of Theorem \ref{theo1a} can be rather conservative, as it does not take account of any possible structural conditions in the matrix $G(q,\theta)$. Additionally the result does not apply to the situation where $U(q,\theta)$ is not full row rank, as in that case it can never be transformed to having a leading diagonal by column operations. This is e.g. the case in Example \ref{example1}. Both structural constraints and possible reduced row rank of $U(q,\theta)$  will be further considered in Section \ref{sect:ident2}.
First we will present some illustrative examples that originate from \cite{weerts_etal_ALCOSP2016}.
\section{Illustrative examples}\label{sec:exam}
\begin{example}[Closed-loop system]
One of the very simple examples to which the results of this paper apply is the situation of a single-loop feedback system, with a disturbance signal on the process output, and a reference input at the process input (controller output), see Figure \ref{fig1}.
\begin{figure}[ht]
\centering{\includegraphics[width=40mm]{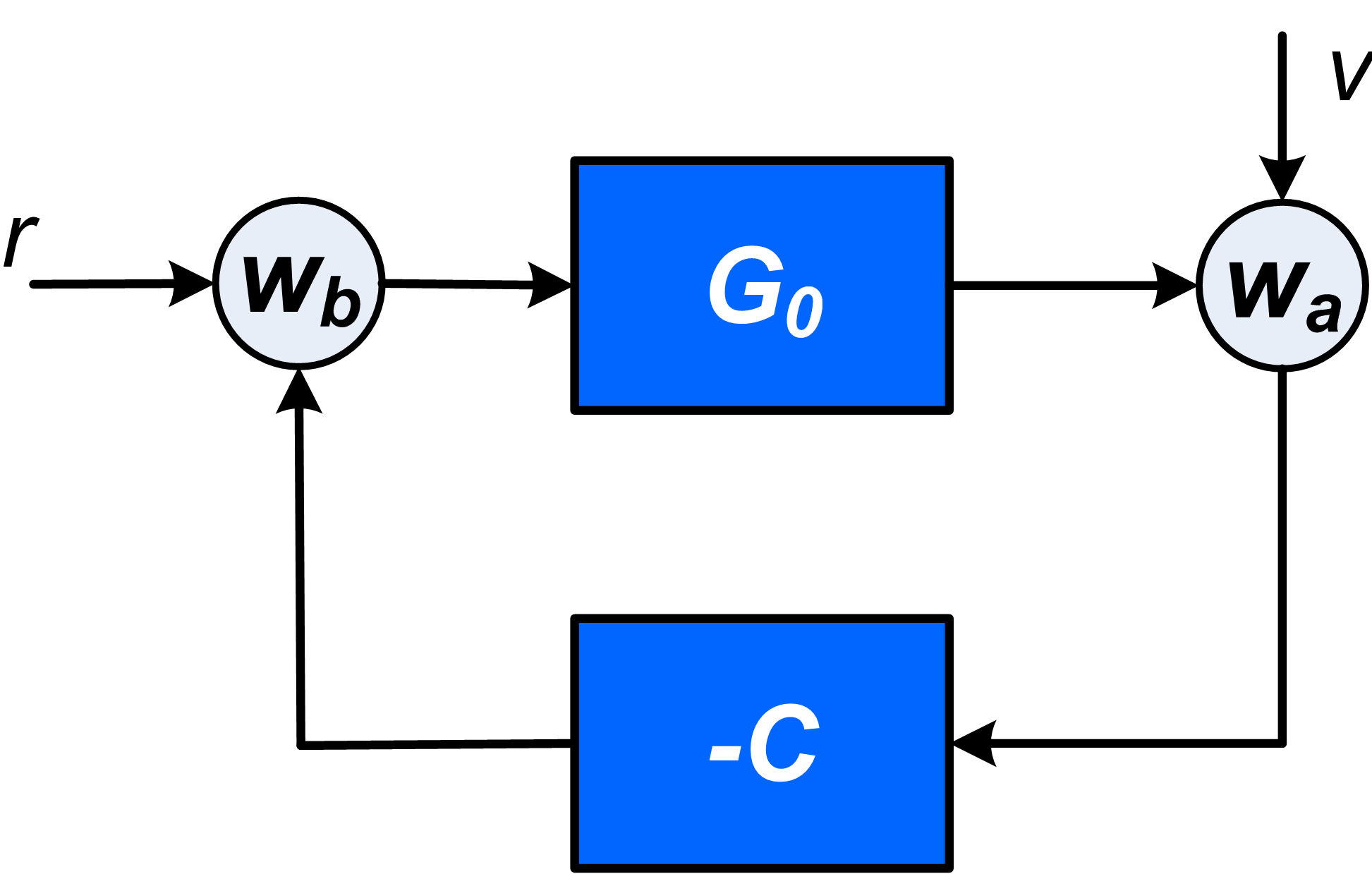}}
\caption{Classical closed-loop configuration.}
\label{fig1}
\end{figure}
The process output $y$ will take the role of node variable $w_a$, while the process input $u$ will be represented by the (noise-free) $w_b$. When parametrizing process $G(q,\theta)$ and controller $C(q,\theta)$, as well as noise model $v(t)=H_a(q,\theta)e(t)$ and the fixed reference filters $R_a(q)=0$, $R_b(q)=1$, it appears that the essential identifiability result of Theorem \ref{theo1a} is reflected by the matrix
\[ U(q,\theta)
 = \begin{bmatrix} H_a(q,\theta) & 0 \\ 0 & 1	\end{bmatrix}. \]
This matrix is square and equal to the diagonal matrix $D$ in the theorem.
Since it is square we have that matrix $F$ will have dimension $2 \times 0$.
The conditions of Theorem \ref{theo1a} are satisfied with $Q=I$, and therefore the closed-loop system is globally network identifiable.
This implies that consistent estimates of $G_0$ and $C$ can be obtained, when identified simultaneously.

In our current setting we consider the simultaneous identification of all modules in the network. In the classical direct method of closed-loop identification, one typically parametrizes the plant model $G$, but not the controller $C$, implying that only part of the network is identified. This can lead to questions of identifiability of part of a network (rather than of the full network). The analysis of such a question can fit into the general setting of Definition \ref{defif} by considering the network property $f(M)=G$, as meant in Remark \ref{remark1}.
\end{example}

\begin{example}[Network example]

In this example we analyze the 5 node network of Figure \ref{fig2} where the noises on nodes $1$ and $2$ are correlated.
\begin{figure}[ht]
\centering{\includegraphics[width=.9\columnwidth]{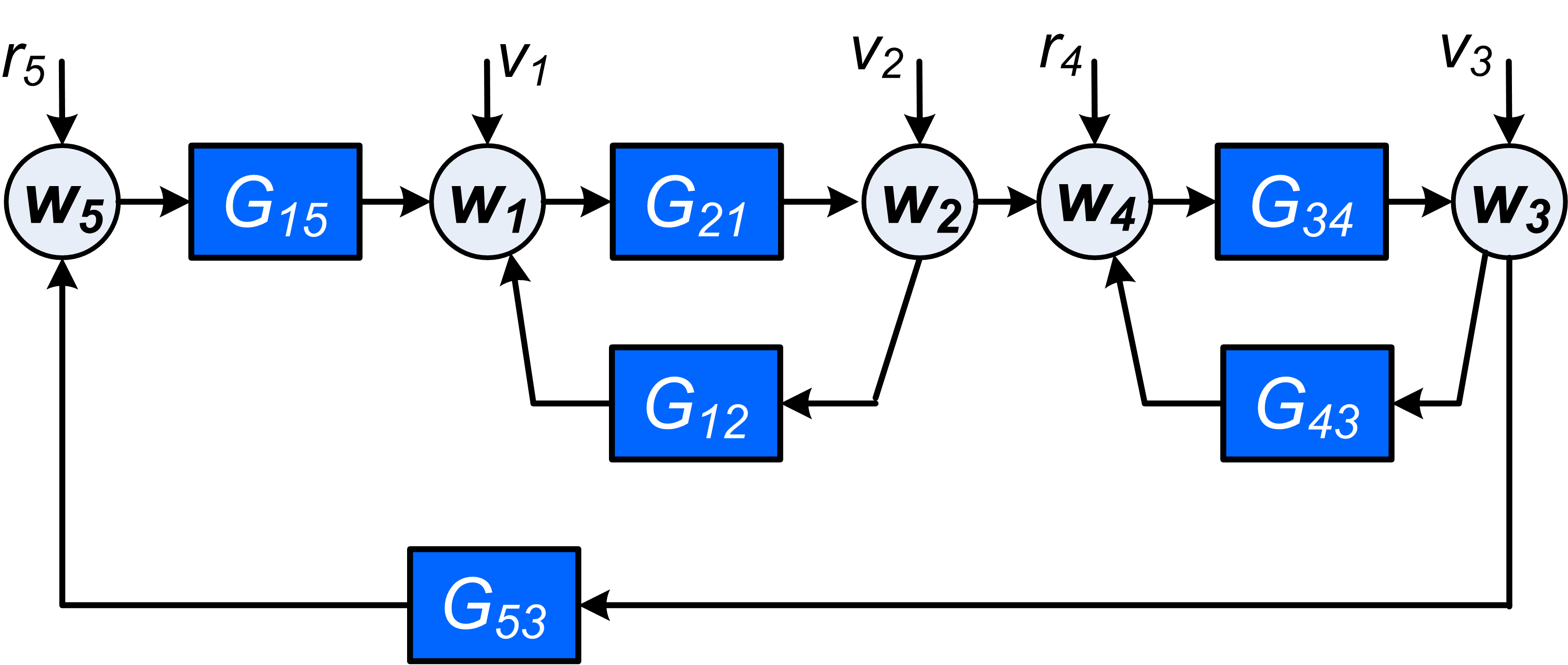}}
\caption{5 node network.}
\label{fig2}
\end{figure}
The nodes are labeled such that the last two are noise-free.
Process noise will be modeled according to
\[ \begin{bmatrix} v_1(t) \\ v_2(t) \\ v_3(t) \end{bmatrix} = \underbrace{
\begin{bmatrix}
	H_{11}(q,\theta) & H_{12}(q,\theta) & 0 \\ H_{21}(q,\theta) & H_{22}(q,\theta) & 0 \\ 0 & 0 & H_{33}(q,\theta)
\end{bmatrix}}_{H_a(q,\theta)}
\begin{bmatrix} e_1(t) \\ e_2(t) \\ e_3(t) \end{bmatrix}. \]
The elements $H_{21}$ and $H_{12}$ are present to allow for modelling correlation between the process noises $v_1$ and $v_2$, while $v_3$ is modelled independently from these two signals. As the external excitation signals $r_4$ and $r_5$ directly affect the two corresponding node signals, without a dynamic transfer, the corresponding $R$ matrices are not parametrized but fixed to $1$. This leads to a matrix $U(q,\theta)$ constructed as
\[
	U(q,\theta) = \begin{bmatrix}
	H_{11}(q,\theta) & H_{12}(q,\theta) & 0 & 0 & 0\\
	H_{21}(q,\theta) & H_{22}(q,\theta) & 0 & 0 & 0 \\
	0 & 0 & H_{33}(q,\theta) & 0 & 0 \\
	0 & 0 & 0 & 1 & 0\\
	0 & 0 & 0 & 0 & 1
\end{bmatrix}.
\]
The condition of Theorem \ref{theo1a} is now checked by attempting to diagonalize the matrix $U(q,\theta)$
by postmultiplication with some filter $Q(q)$ which does not depend on $\theta$.
Due to the correlated noise it is not possible to diagonalize the matrix in this way.
Note that by adding external excitations to nodes $1$ and $2$, leading to the addition of fixed unit vector columns in $U(q,\theta)$, we can make the model set globally network identifiable.
\end{example}
\section{Network identifiability in case of structure restrictions}  \label{sect:ident2}
When there are structure restrictions in $G(q,\theta)$ or matrix $U(q,\theta)$ is not full row rank, as in Example \ref{example1}, the result of Theorem \ref{theo1a} is conservative and/or even does not apply. Structure restrictions in $G(q,\theta)$ are typically represented by fixing some modules, possibly to $0$, on the basis of assumed prior knowledge.
For these cases of structure restrictions, in \citep{Goncalves08} necessary and sufficient conditions have been formulated for satisfying (\ref{equivTl}) at a particular model $M_0$. The conditions are formulated in terms of nullspaces that can  not be checked without knowledge of the underlying network. Since we are most interested in global identifiability of a full model set, rather than in a particular model, we will further elaborate and generalize these conditions and present them in a form where these conditions can \red{effectively be checked.}

First we need to introduce some notation.\\
In line with the reasoning in Section \ref{sec:alglp}, we suppose that each row $i$ of $G(\theta)$, has $\alpha_i$ parameterized transfer functions, and row $i$ of $U(\theta)$ has $\beta_i$ parametrized transfer functions, and we define
the $L \times L$ permutation matrix $P_i$, and the $(K+p)\times (K+p)$ permutation matrix $Q_i$,
such that all parametrized entries in the considered row  of $(I-G(q,\theta))\red{P_i}$ are \red{gathered on the left hand side}, and all parametrized entries in the considered row of $U(q,\theta)\red{Q_i}$ are \red{gathered on the right hand side}, i.e.
\begin{eqnarray}
	 	(I-G(\theta))_{i \star} P_i & = & \begin{bmatrix} (I-G(\theta))_{i \star}^{(1)} & (I-G)_{i \star}^{(2)}	 \end{bmatrix} \label{eq:pit} \\
U(\theta)_{i\star}Q_i & = & \begin{bmatrix} U_{i\star}^{(1)} & U(\theta)_{i\star}^{(2)} \end{bmatrix} \label{eq:qit}
\end{eqnarray}
Next we define the transfer matrix $\check T_i(q,\theta)$ of dimension $\alpha_i \times (K+p-\beta_i)$, as the submatrix of $T(q,\theta)$ that is constructed by taking the row numbers that correspond to the columns of $G(q,\theta)_{i\star}$ that are parametrized, and by taking the column numbers that correspond to the columns of $U(q,\theta)$ that are not parametrized. This is formalized by
\begin{equation}
\label{eq:Ti}
	\check T_i(q,\theta) := \left[ I_{\alpha_i}\ \ 0 \right] P_i^{-1} T(q,\theta)Q_i\begin{bmatrix}I_{K+p-\beta_i} \\ 0  \end{bmatrix}.
\end{equation}
The following Theorem now specifies necessary and sufficient conditions for the central  identifiability condition (\ref{equivTl}).

\begin{theorem} \label{thm:theo3}
Let $\M$ be a network model set for which the conditions of one of the Propositions \ref{propT}-\ref{propT3} are satisfied, and that additionally satisfies the following properties:
\begin{itemize}
\item[a.] Every parametrized entry in the model $\{ M(z,\theta), \theta\in\Theta\}$ covers the set of all proper rational transfer functions;
\item[b.] All parametrized transfer functions in the model $M(z,\theta)$ are parametrized independently (i.e. there are no common parameters).
\end{itemize}
Then
	\begin{enumerate}
		\item $\M$ is globally network identifiable at $M(\theta_0)$ if and only if
		\begin{itemize}
			\item 	
			each row $i$ of the transfer function matrix $\begin{bmatrix} G(\theta) & U(\theta)	 \end{bmatrix}$ has at most $K+p$ parameterized entries, and
			\item
			for each $i$, $\check T_i(\theta_0)$ defined by (\ref{eq:Ti}) has full row rank.
		\end{itemize}
		\item $\M$ is globally network identifiable if and only if
		\begin{itemize}
			\item
			each row $i$ of the transfer function matrix $\begin{bmatrix} G(\theta) & U(\theta)	 \end{bmatrix}$ has at most $K+p$ parameterized entries, and
			\item
			for each $i$, $\check T_i(\theta)$ defined by (\ref{eq:Ti}) has full row rank for all $\theta \in \Theta$.	\hfill $\Box$		
		\end{itemize}
	\end{enumerate}	
\end{theorem}
The proof is provided in the Appendix.
The condition on the maximum number of parametrized entries in the transfer function matrix, reflects a condition that the number of parametrized transfers that map into a particular node, should not exceed the total number of excitation signals plus white noise signals that drive the network.
The check on the row rank of matrices $\check T_i$ is an explicit way to check the related nullspace condition in \citep{Goncalves08}. The assumption (a.) in the Theorem, refers to the situation that we do not restrict the model class to any finite dimensional structure, but that we consider the situation that could be represented by a non-parametric identification of all module elements.
\begin{remark}
The condition on the maximum number of parametrized entries per row in the parametrized matrix seems closely related to a similar condition for structural identifiability of (polynomial) ARMAX systems, as formulated in Theorem 2.7.1. of \cite{Hannan&Deistler:88}. A further analysis of this relationship is beyond the scope of this paper, and will be explored elsewhere.
\end{remark}
\red{
\begin{remark}
The identifiability results as formulated in the above Theorem can also be applied row-wise to the composed matrix $\begin{bmatrix}  G(\theta) & U(\theta)\end{bmatrix}$. This implies that the elements of row $i$ of this matrix are uniquely identifiable, if the formulated conditions are satisfied for the particular value of $i$ only. This fits with the reasoning in Remark \ref{remark1}, and can be simply verified in the proof of the Theorem. This aspect is also addressed in \cite{weerts_etal_2015}.
\end{remark}
}
The results of this Section can be applied to Example \ref{example1}.

\begin{example}[Example \ref{example1} continued]
In Example \ref{example1} a model set has been defined with $U=R$ not full row rank, and hence Theorem \ref{theo1a} is not suitable for checking its network identifiability.
Now with the introduction of necessary and sufficient conditions in Theorem \ref{thm:theo3} we can evaluate the network identifiability property of the model set in Example \ref{example1} easily.
Consider a model set with $G$ and $R$ as defined in (\ref{eq:gr}), without noise model (i.e. $p=0$) such that $U=R$, and satisfying assumptions (a.) and (b.) of Theorem \ref{thm:theo3}.
Global network identifiability at $\mathcal S_1$ and $\mathcal S_2$ is evaluated by checking the two conditions of Theorem \ref{thm:theo3}.
First it is easily verified that $\begin{bmatrix} G(\theta) & U \end{bmatrix}$ has at most $2=K+p$ parameterized transfer functions on each row. The second condition is checked by evaluating the rank of the appropriate sub-matrices defined in
(\ref{eq:Ti}) on the basis of the $T$-matrices for $\mathcal S_1$ and $\mathcal S_2$ given in (\ref{eq:t12}).\\
For $\mathcal S_1$ we need to check the conditions for rows 1-3 accordingly. Then by considering (\ref{eq:ex_mod_sys1}) we can determine $\check T_i (q,\theta_1)$ as appropriate submatrices of $T(q,\theta_1)$. For all rows $i$, $Q_i=I$, since $U$ is not parametrized, and so we need to consider all columns of $T(q,\theta_1)$. For $i=1$, $\check T_i(q,\theta_1)$ is defined by selecting the second and third row of $T(q,\theta_1)$, corresponding with the columns of parametrized elements in $G_{1\star}(q,\theta_1)$, i.e.
\[  \check T_1(q,\theta_1) = \begin{bmatrix} A & 1 \\ AB+1 & B  \end{bmatrix}; \]
while for $i=2$ we need to select rows one and three, and for $i=3$ rows one and two,
corresponding with the columns of parametrized elements in $G_{2\star}(q,\theta_1)$ and
$G_{3\star}(q,\theta_1)$, respectively, leading to %
\[ \check T_2(q,\theta_1) = \begin{bmatrix} 1 & 0 \\ AB+1 & B\end{bmatrix},\ \ \ \
 \check T_3(q,\theta_1) = \begin{bmatrix} 1 & 0 \\ A & 1     \end{bmatrix}. \]
Since all three $\check T$-matrices are full row-rank, the conditions for global network identifiability at $\mathcal S_1$ are satisfied which verifies the conclusion of Example \ref{example1}. \\
For $\mathcal S_2$ a similar check needs to done on the basis of (\ref{eq:ex_mod_sys2}), leading to
\[ \check T_2(q,\theta_2) = \begin{bmatrix} 1 & 0 \\ A+1 & 0  \end{bmatrix} \]
which obviously does not have full rank, confirming that the model set is not globally network identifiable at $\mathcal S_2$.\\
If we would restrict the model set to satisfy $G_{21}(\theta)=0$, it can simply be verified that the conditions for global network identifiability at $\mathcal{S}_2$ are satisfied, which is confirmed by the analysis in (\ref{eq:ex_uonunique}).
\end{example}
When information about the 'true' network is used then one can obtain results that allow us to distinguish between certain networks. However we are mainly interested in results that allow us to distinguish between all networks in a model set, since in an identification setting the true network structure/dynamics will not be known.
\section{Discussion on signal ordering assumption}
\label{sec:disc}
In Assumption \ref{asspi} we have formulated a condition on an ordering property of the model set. In this Section we will further discuss this Assumption and how it can be dealt with.

Our definitions of models and model sets in Section \ref{sect:pred} only consider models that have the ordering property. So, for discussing the situation of models that do not have this property, we need to slightly adapt our definition.
\begin{definition}[Network model without ordering]
\label{def3}
A network model without ordering property is defined by the quadruple
\begin{equation}
\label{eq:mm}
 M = (G,R,\tilde H,\tilde\Lambda)
\end{equation}
with $\tilde H \in \R^{L\times L}(z)$ monic, and $\tilde\Lambda \in \R^{L\times L}$, and $G$ and $R$ as defined before in Definition \ref{def1}.
\end{definition}
%
First of all, if we are considering network identifiability at a particular (unordered) model $M_0=(G_0,R_0,\tilde H_0,\tilde\Lambda_0)$, then the covariance matrix $\tilde\Lambda_0$ carries the information of the rank $p$ as well as the information for re-ordering the node signals $w$ in such way that, after reordering, the model satisfies the ordering property of Assumption 1. This can be understood by realizing that rank $\tilde\Lambda_0 = p$, and that there exists a permutation matrix $\Pi$ such that $[I_p\ 0]\Pi^T\tilde\Lambda_0\Pi [I_p\ 0]^T = \Lambda_0$, the rank-$p$ covariance matrix of the ordered model. That same permutation matrix can then be applied to $w$, to reorder the node signals in the model so as to arrive at its ordered equivalent. So when addressing the problem of global identifiability at a particular model, the model information intrinsically contains the information how to order the signals to satisfy the ordering property.

In more general situations, the required information for determining $p$ and for reordering the node signals can be retrieved from data, $T_{wr}$ and $\Phi_{\bar v}(\omega)$. In particular we can observe that on the basis of
\[ \bar v = (I-G)^{-1}v \]
and invertibility of $(I-G)$, it is clear that rank $\Phi_{\bar v} = \mbox{ rank } \Phi_v = p$, and more specifically, by using the monicity property of $\tilde H$, that rank $\Phi_{\bar v}^\infty =  \mbox{ rank } \tilde \Lambda = p$. So for a particular model $M(\theta_0)$, $p$ can be obtained directly from $\Phi_{\bar v}^\infty(\theta_0)$. \\
A similar situation occurs for the ordering of signals as assumed in Assumption \ref{asspi}, as is formulated next.
\begin{proposition}
\label{propord}
Consider a network model $M_0=M(\theta_0)$ according to Definition \ref{def3}, with rank $\Phi_{\bar v}(\theta_0) = p$. If either one of the following conditions is satisfied:
\begin{enumerate}
\item $G^\infty(\theta_0) = 0$;
\item $G^\infty(\theta_0)$ has a known pattern of $0$'s, that guarantees that there are no algebraic loops, and $\Phi_v^\infty(\theta_0)$ is diagonal;
\item Each row of $\left[G^\infty(\theta_0)\ R^\infty(\theta_0)\right]$ has at most $K$ nonzero elements, and \\ for each $i=1,\cdots L$, the matrix $\check T_i^\infty(\theta_0)$ (\ref{eq:tinf}) has full row rank,
\end{enumerate}
then on the basis of $T_{wr}^\infty(\theta_0)$ and $\Phi_{\bar v}^\infty(\theta_0)$ a permutation matrix $\Pi$ can be constructed that reorders the node signals $w$ in such a way that the permuted model satisfies the ordering property as meant in Assumption \ref{asspi}.
\end{proposition}
A proof is collected in the Appendix. The reasoning that underlies this result, is that under the formulated conditions the covariance matrix $\tilde\Lambda_0$ can be uniquely retrieved from the data. And based on $\tilde\Lambda_0$ a permutation matrix can then be found that reorders the node signals into a (reordered) model that satisfies the ordering property.\\
The conditions of this Proposition are basically the same as the ones applied in Propositions \ref{propT}, \ref{propT2} and \ref{propT3} for analyzing identifiability.

The results in this section show that the ordering property of Assumption 1 is not a restriction if we consider the identifiability of a model set at a particular model (local analysis). This is due to the fact that in that particular model, either the model information or the measurement data in the form of $T_{wr}^\infty$ and $\Phi_{\bar v}^\infty$ carry enough information to find a permutation matrix to arrive at a permuted model that does satisfy Assumption 1.

\section{Conclusions}
	The objective of this paper has been to obtain conditions on the presence and location of excitation and disturbance signals and conditions on the parameterized model set such that a unique representation of the full network can be obtained.
	A property called global network identifiability has been defined to ensure this unique representation, and results have been derived to analyze this property for the case of dynamic networks allowing correlated noises on node signals, as well as rank-reduced noise.
Three key ingredients for a network identifiable model set are: presence and location of external excitation signals, modeled correlations between disturbances, and prior (structural) knowledge on the network that is incorporated in the model set.

\section{Acknowledgement}
The authors gratefully acknowledge discussions with Michel Gevers and Manfred Deistler on the topic and presentation of this paper, and Manfred Deistler in particular for the suggestion of using signal spectra as a basis of model equivalence in identifiability.

\bigskip
\appendix
{\bf Appendix}
\def\thetheorem{A.\arabic{theorem}}
\def\thelemma{A.\arabic{lemma}}
\def\theremark{A.\arabic{remark}}
\def\theproposition{A.\arabic{proposition}}
\setcounter{lemma}{0}
\setcounter{remark}{0}
\setcounter{proposition}{0}
%

\vspace*{-3mm}
\section{Proof of Proposition \ref{propT}}
Since in the considered situation
\[ T_{we}(\theta) := (I-G(\theta))^{-1}H(\theta) \]
has an upper $p \times p$ part which is monic, while
\begin{equation} \label{eqv1}
 \Phi_{\bar v}(\theta) = T_{we}(\theta)\Lambda(\theta)T_{we}(\theta)^*
\end{equation}
it follows that (\ref{eqv1}) satisfies the conditions of the unique spectral factorization in Lemma \ref{lemrr}a, if $p<L$. If $p=L$ it satisfies the conditions of the standard spectral factorization.   Therefore $T_{we}$ and $\Lambda$ are uniquely determined by $\Phi_{\bar v}$, or in other words
\[ \{\Phi_{\bar v}(\theta_1) = \Phi_{\bar v}(\theta_0)\} \Longrightarrow
\left\{ \begin{array}{c}  T_{we}(\theta_1) = T_{we}(\theta_0) \\ \Lambda(\theta_1) = \Lambda(\theta_0)   \end{array} \right. .
\]
Since $T_{wr}(\theta_1)=T_{wr}(\theta_0)$ is in the premise of (\ref{equivTP}) and $\Lambda(\theta_1)=\Lambda(\theta_0)$ is implied by the premise of the equality of the spectra, as indicated above, the result follows directly. \hfill $\Box$
\section{Proof of Proposition \ref{propT2}}
First we treat the full-rank situation that $p=L$.

In this situation
\[ \Phi_{\bar v}(z,\theta) := (I-G(\theta))^{-1}H(\theta)\Lambda(\theta) H(\theta)^*(I-G(\theta))^{-*} \]
and using the property that $H$ is monic leads to
\[ \Phi_{\bar v}^\infty(\theta)\! :=\!\lim_{z\rightarrow \infty}\! \Phi_{\bar v}(z,\theta) =
        (I\!-\!G^\infty(\theta))^{-1}\Lambda(\theta)(I\!-\!G^\infty(\theta))^{-T}. \]
The algebraic loop condition now implies that $\Pi^T(I-G^\infty(\theta))^{-1}\Pi$ is upper unitriangular\footnote{upper unitriangular is upper triangular with $1$'s on the diagonal.} and
(leaving out arguments $\theta$ for brevity):
\begin{eqnarray*}
\lefteqn{\Pi^T \Phi_{\bar v}^\infty \Pi =} \\
& & \underbrace{\Pi^T(I-G^\infty)^{-1}\Pi}_{L}\cdot \underbrace{\Pi^T\Lambda\Pi}_{D} \cdot \underbrace{\Pi^T(I-G^\infty)^{-T}\Pi}_{L^T}.
\end{eqnarray*}
With $D$ being diagonal and $L$ upper unitriangular, this represents a unique $LDL^T$ decomposition of the permuted spectrum. As a result $\Lambda$ is uniquely determined from $\Phi_{\bar v}$.

Spectral factorization of $\Phi_{\bar v}$ leads to a unique decomposition
\[ \Phi_{\bar v} = \hat H \hat\Lambda \hat H^* \]
with $\hat H$ monic, stable and minimum-phase, but $\hat \Lambda$ not necessarily diagonal. Since $\hat \Lambda$ is full rank, there is a nonsingular matrix $B$ such that $\hat\Lambda = B\Lambda B^T$, leading to the unique spectral decomposition:
\[ \Phi_{\bar v} = \hat HB\Lambda B^T\hat H^*,  \]
where $\hat H B = T_{we}$. As a result, $T_{we}$ is uniquely determined from $\Phi_{\bar v}$, and the proof follows along the same steps as in the proof of Proposition \ref{propT}.

\medskip
Now we turn to the situation $p < L$.

When applying the spectral decomposition of Lemma \ref{lemrr}b to $\Phi_v$ it follows that
\[ \Phi_{\bar v}(z,\theta) = (I-G(\theta))^{-1}\breve H(\theta)\breve \Lambda(\theta) \breve H(\theta)^*(I-G(\theta))^{-*} \]
with $\breve H$ square and monic, and structured according to
\[ \breve H = \begin{bmatrix} H_a & 0 \\ H_b - \Gamma & I \end{bmatrix},\ \ \mbox{and }
\breve\Lambda = \begin{bmatrix} I \\ \Gamma \end{bmatrix} \Lambda \begin{bmatrix} I \\ \Gamma \end{bmatrix}^T. \]
Since by assumption
$\Phi_v^\infty$ is diagonal, it follows that
$\Gamma := \lim_{z\rightarrow\infty}H_b(z) = 0$ and
\[ \breve\Lambda = \begin{bmatrix} \Lambda & 0 \\ 0 & 0 \end{bmatrix}. \]
As a result
\[ \Phi_{\bar v}^\infty = (I-G^\infty(\theta))^{-1}\breve\Lambda(\theta)(I-G^\infty(\theta))^{-T} \]
with $\breve\Lambda(\theta)$ diagonal. Then exactly the same reasoning as above with a permutation of the
signals to turn $(I-G^\infty)^{-1}$ into a unitriangular matrix, shows that $\breve\Lambda$ and therefore also $\Lambda$ is uniquely determined from $\Phi_{\bar v}$.

With $\Lambda$ known, the decomposition $\Phi_{\bar v} = T_{we}\Lambda T_{we}^*$ uniquely determines $T_{we}$ from $\Phi_{\bar v}$. The proof then follows the same same steps as in the proof of Proposition \ref{propT}. \hfill $\Box$

\section{Proof of Proposition \ref{propT3}}
This proof consists of 2 steps. The first step is to use $T_{wr}$ to uniquely determine the feedthrough of $G$, i.e.
\begin{equation}
	T_{wr}^\infty (\theta_1)= 	T_{wr}^\infty (\theta_0)
	\Rightarrow
	G^\infty (\theta_1) = G^\infty (\theta_0).
\end{equation}
The left hand side of the above implication can be written as
\begin{equation} \label{eq:gtinf}
	(I-G^\infty (\theta_0)) T_{wr}^\infty (\theta_1)= 	R^\infty(\theta_0).
\end{equation}
Consider row $i$ of the matrix equation (\ref{eq:gtinf}), and apply the following reasoning for
each row separately.
By inserting the permutation matrices $P_i$ and $Q_i$, defined in (\ref{eq:pi})-(\ref{eq:qi}), we obtain for
row $i$:
\begin{equation}
	(I-G^\infty (\theta_0)) P_i^{-1} P_i T_{wr}^\infty (\theta_1) Q_i= 	R^\infty(\theta_0) Q_i
\end{equation}
leading to
\begin{eqnarray}
\lefteqn{	(I-G^\infty(\theta_0))^{(1)}_{i\star} T^{(1)}_i(\theta_1)
	+ (I-G^\infty)^{(2)}_{i\star}	T^{(2)}_i(\theta_1)
	= } \nonumber \\
& & = \begin{bmatrix} {R^\infty}^{(1)}_{i\star} & {R^\infty}^{(2)}_{i\star}(\theta_0)	 \end{bmatrix},\label{eq:veceq}
\end{eqnarray}
with $P_i T_{wr}^\infty (\theta_1) Q_i = \begin{bmatrix}T^{(1)}_i(\theta_1) \\ T^{(2)}_i(\theta_1) \end{bmatrix}$.
Note that, as defined by (\ref{eq:tinf}), $\check T_i^\infty(\theta) = T^{(1)}_i (\theta) \begin{bmatrix} I_{K-\beta_i} \\ 0 \end{bmatrix}$.

When considering the left $1 \times (K - \beta_i)$ block of the vector equation (\ref{eq:veceq}), while using the expression for $\check T_i^\infty(\theta)$ above, we can write
\begin{equation}
	(I-G^\infty(\theta_0))^{(1)}_{i\star} \check T_i^\infty(\theta_1)
	+ \rho(\theta_1)
	=
	{R^\infty}^{(1)}_{i\star},
\end{equation}
with $\rho(\theta_1)$ the left $1 \times (K-\beta_i)$ block of $(I-G^\infty)^{(2)}_{i\star}	 T^{(2)}_i(\theta_1)$.
Now $\rho(\theta_1)$ and ${R^\infty}^{(1)}_{i\star}$ are independent of parameter $\theta_0$, which implies that, if $\check T_i^\infty(\theta_1)$ has full row rank, then all the parametrized elements in $(I-G^\infty(\theta_0))_{i\star}$ are uniquely determined.

Then the second step is to determine $\Lambda$ and $T_{we}$.
By writing the spectrum of $\bar v$ as
\[
	\Phi_{\bar v}^\infty = (I-G^\infty)^{-1} H^\infty(\theta) \Lambda(\theta) (H^\infty(\theta))^T (I-G^\infty)^{-T}
\]
we obtain through pre- and post-multiplication:
\[
(I-G^\infty) \Phi_{\bar v}^\infty (I-G^\infty)^{T}=	
\begin{bmatrix}
		\Lambda(\theta) & \Lambda(\theta) \Gamma^T(\theta)
		\\
		\Gamma(\theta) \Lambda(\theta) & \Gamma (\theta) \Lambda(\theta) \Gamma^T(\theta)
	\end{bmatrix}
\]
where $\Gamma := \lim_{z \rightarrow \infty} H_b(z,\theta)$. For given $G^\infty$ (from step 1), and
given $\Phi_{\bar v}$, this equation provides a unique $\Lambda$, such that $T_{we}$ can be uniquely obtained from
\begin{equation}
	\Phi_{\bar v} = T_{we}(\theta) \Lambda T_{we}^*(\theta).
\end{equation}
The proof then follows the same steps as the proof of Proposition 1.
\hfill $\Box$

\section{Proof of Theorem \ref{theo1a}}
a) It will be shown that under the condition of the theorem, the equality $T(q,\theta) = T(q,\theta_0)$ implies $M(\theta)=M(\theta_0)$ for all $\theta \in \Theta$. With the definition of $\Theta_0$, the equality of the $T$-matrices implies that we can restrict to $\theta \in \Theta_0$. That same equality induces
\begin{equation}
(I-G(\theta))^{-1}U(\theta) = (I-G(\theta_0))^{-1}U(\theta_0) \label{thm1eq1}
\end{equation}
and postmultiplication with $Q$ leads to
\[(I\! -\! G(\theta))^{-1} \begin{bmatrix} D(\theta) & F(\theta) \end{bmatrix}\! =\! (I\! -\! G(\theta_0))^{-1} \begin{bmatrix} D(\theta_0) & F(\theta_0) \end{bmatrix}, \]
with $D(\theta)$ diagonal and full rank for all $\theta \in \Theta_0$.\\
The left square $L \times L$ blocks in both sides of the equation can now be inverted to deliver
$ D(\theta)^{-1} (I-G(\theta))  = D(\theta_0)^{-1} (I-G(\theta_0)). $
	Due to zeros on the diagonal of $G(\theta)$ and $G(\theta_0)$ and the diagonal structure of $D(\theta)$ and $D(\theta_0)$, it follows that $D(\theta)=D(\theta_0)$ and consequently $G(\theta)=G(\theta_0)$. Then by (\ref{thm1eq1}) it follows that $U(\theta)=U(\theta_0)$ and $M(\theta)=M(\theta_0)$.\\
b) For part (b) it needs to be shown that the implication under (a) holds true for {\it any} $M(\theta_0)$ in $\M$. It is direct that this is true, following a similar reasoning as above, if we extend the parameter set to be considered from $\Theta_0$ to $\Theta$.
\hfill $\Box$

\section{Proof of Theorem \ref{thm:theo3}}

We will first provide the proof for situation (1).\\
The left hand side of the implication (\ref{equivT}) can be written as
	\begin{equation} \label{eq:row}
		(I-G(\theta)) T = U(\theta),
	\end{equation}
	where we use shorthand notation $T = T(\theta_0)$, $G(\theta) = G(\theta_1)$ and $U(\theta)=U(\theta_1)$.
	Consider row $i$ of the matrix equation (\ref{eq:row}), and apply the following reasoning for each row separately.
    By inserting the permutation matrices $P_i$ and $Q_i$, defined in (\ref{eq:pit}),(\ref{eq:qit}) we obtain for row $i$:
\begin{equation}
(I-G(\theta))_{i\star} P_iP_i^{-1}TQ_i = U_{i \star}(\theta)Q_i
\end{equation}
leading to
	\begin{equation}\label{eq11}
		(I-G(\theta))_{i\star}^{(1)} T^{(1)}_i + (I-G)_{i \star}^{(2)} T^{(2)}_i = \begin{bmatrix} U_{i \star}^{(1)} &  U(\theta)_{i\star}^{(2)} \end{bmatrix},
	\end{equation}
with $P_i^{-1}TQ_i = \begin{bmatrix} T^{(1)}_i \\ T^{(2)}_i \end{bmatrix}$.
Note that $\check T_i =  T_i^{(1)}\begin{bmatrix} I_{K+p-\beta} \\ 0 \end{bmatrix}$.

\textbf{Sufficiency:}
	\\
When considering the left $1\times (K+p-\beta_i)$ block of the vector equation (\ref{eq11}), while using the expression for $\check T_i$ above, we can write
\begin{equation}\label{eq:perm}
(I-G(\theta))_{1\star}^{(1)}\breve T_i + \rho = U_{i \star}^{(1)},
\end{equation}	
with $\rho$ the left $1\times (K+p-\beta_i)$ block of $(I-G)_{i\star}^{(2)}T_i^{(2)}$. \\
Now $\rho$ and $U_{i \star}^{(1)}$ are independent of $\theta$, which implies that, if $\check T_i$ has full row rank, then all the parametrized elements in $(I-G(\theta))_{i\star}$ are uniquely determined. Then through (\ref{eq11}) the parametrized elements in $U_{i\star}(\theta)$ are also uniquely determined.
	
By assumption we know that one solution to (\ref{eq:row}) is given by $G(\theta_0)$ and $U(\theta_0)$.
Since the solution is unique, and $G(\theta_0)$ and $U(\theta_0)$ are a possible solution we know that $G(\theta_0)$ and $U(\theta_0)$ must be the only solution. This proves the validity of the implication (\ref{equivT}).

\textbf{Necessity of condition 2:}
	\\
	If the matrix $\check T_i(\theta_0)$ is not full row rank, then it has a non-trivial left nullspace. Let the rational transfer matrix $X \neq 0$ of dimension $1 \times \alpha_i$ be in the left nullspace of $\check T_i$. 	Then there also exists a proper, rational and stable $X_p$ in the left nullspace of $\check T_i$.
Then (\ref{eq:perm}) can also be written as 	
	\begin{equation}
		\left( (I-G(\theta))_{i\star}^{(1)} + X_p \right) \check T_i + \rho = U_{i \star}^{(1)}.
	\end{equation}
	By the formulated assumptions (a) and (b) it holds that each parameterized transfer function can be any proper rational transfer function, and that these parameterized transfer functions do not share any parameters.
	This implies that $G(\theta_1)_{i\star} \in \M$ and $ (G(\theta_1)_{i\star} - X_p ) \in \M$ refer to two different model rows of $G$ in the model set, that generate the same network transfer function $T$.
	Hence implication (\ref{equivT}) can not hold. \hfill $\Box$

	\textbf{Necessity of condition 1:}
	\\
If $\alpha_i + \beta_i > K+p$, then $\check T_i(\theta_0)$ will be a tall matrix which can never have a full row rank. Then because of the necessity of the row rank condition on $\check T_i(\theta_0)$, necessity of condition 1 follows immediately. \hfill $\Box$
	
		\textbf{Proof of situation (2): For all $\theta \in \Theta$:}\\
	For every $\theta \in \Theta	$ we can construct $T(\theta)$ with related $\check T_i(\theta)$ of full row rank, and the reasoning as presented before fully applies.
	If for some $\theta \in \Theta$ we can not construct this full row rank $\check T_i(\theta)$ there exists a model in the model set which is not identifiable, and hence the model set is not globally network identifiable in $\M$. \hfill $\Box$
%
\section{Proof of Proposition \ref{propord}}
The expression for $\Phi_{\bar v}$ is given by (discarding arguments $\theta_0$):
\begin{equation}
\label{eq:ph}
\Phi_{\bar v} =  [I-G]^{-1}\tilde H \tilde\Lambda \tilde H^*[I-G]^{-*}.
\end{equation}
while $T_{wr} = [I-G]^{-1}R$.
Because $\tilde H$ is monic, the expression for $\Phi_{\bar v}^\infty$ reduces to:
\begin{equation}
\label{eq:poo}
\Phi_{\bar v}^\infty = [I-G^\infty]^{-1} \tilde\Lambda [I-G]^{-*}.
\end{equation}
We are now going to show that under the different conditions, $\tilde \Lambda$ can be uniquely derived from $\Phi_{\bar v}^\infty$ and $T_{wr}^\infty$.

{\bf Situation of strictly proper modules (Proposition \ref{propT}).} \\
Since we know that $G^\infty=0$ it follows immediately from (\ref{eq:poo}) that $\Phi_{\bar v}^\infty  = \tilde \Lambda$, showing that $\tilde \Lambda$ can be directly obtained from $\Phi_{\bar v}^\infty$.

{\bf Situation of diagonal $\Lambda$ and no algebraic loops (Proposition \ref{propT2}).} \\
If $\Phi_v^\infty$ is diagonal then also $\tilde \Lambda$ is diagonal. We consider (\ref{eq:poo}). Based on the algebraic loop condition, we can construct
a permutation matrix $\Pi$ such that $\Pi^T(I-G^\infty)^{-1}\Pi$ is upper unitriangular.
Then:
\begin{eqnarray*}
\lefteqn{\Pi^T \Phi_{\bar v}^\infty \Pi =} \\
& & \underbrace{\Pi^T(I-G^\infty)^{-1}\Pi}_{L}\cdot \underbrace{\Pi^T\tilde\Lambda\Pi}_{D} \cdot \underbrace{\Pi^T(I-G^\infty)^{-T}\Pi}_{L^T}.
\end{eqnarray*}
With $D$ being diagonal and $L$ upper unitriangular, this represents a unique $LDL^T$ decomposition of the permuted spectrum. As a result $\tilde\Lambda$ is uniquely determined from $\Phi_{\bar v}^\infty$.

{\bf Situation of algebraic loops (Proposition \ref{propT3}).} \\
The proof of Proposition \ref{propT3} shows that under the given conditions, $G^\infty$ is uniquely determined from $T_{wr}$.
Then (\ref{eq:poo}) leads to the expression
\begin{equation}
[I-G^\infty]\Phi_{\bar v}^\infty [I-G^\infty]^{*}  =  \tilde\Lambda.
\end{equation}
showing that $\tilde \Lambda$ can be uniquely determined.

In all three situations considered, the matrix $\tilde\Lambda$ is uniquely determined from $\Phi_{\bar v}^\infty$ and possibly $T_{wr}^\infty$. Then there exists a permutation matrix $\Pi$ that reorders the signals $v$ in such a way that $\Pi \tilde \Lambda\Pi^T$ is a matrix of which the left upper $p\times p$ part is full rank. If we apply this reordering of signals, determined by $\Pi$, to the node signals $w$, then we arrive at a permuted model that has the ordering property, according to Assumption \ref{asspi}.

\bibliographystyle{plainnat}		
\bibliography{IdentifiabilityLibrary}

\end{document}